\documentclass[conference]{IEEEtran}
\IEEEoverridecommandlockouts
\usepackage{cite}
\usepackage{amsmath,amssymb,amsfonts}
\usepackage{multicol}
\usepackage{algorithmic}
\usepackage{graphicx}
\usepackage{textcomp}
\usepackage{xcolor}
\usepackage{definice}

\usepackage{siunitx}

\newcounter{steps}
\newenvironment{descriptionA}
{\begin{list}{\textbf{Step} \textbf{\arabic{steps}}:}%
		{\setlength\labelsep{5pt}%
			\setlength\itemindent{5pt}%
			\setlength\parsep{0pt}%
			\setlength\leftmargin{0pt}%
			\setlength\labelwidth{0pt}%
			\usecounter{steps}}}%
{\end{list}}

\def\real{\mathbb{R}}

\def\bfD{\mathbf D}
\def\bfF{\mathbf F}
\def\bfG{\mathbf G}

\def\bfK{\mathbf K}

\def\bfQ{\mathbf Q}
\def\bfP{\mathbf P}
\def\bfR{\mathbf R}
\def\bfS{\mathbf S}

\def\bfg{\mathbf g}
\def\bfh{\mathbf h}

\def\bfv{\mathbf v}
\def\bfw{\mathbf w}
\def\bfx{\mathbf x}
\def\bfy{\mathbf y}
\def\bfz{\mathbf z}
\def\bfnul{\mathbf 0}

\def\calN{\mathcal{N}}
\def\calO{\mathcal{O}}

\def\mean{\mathsf{E}}

\def\cov{\mathsf{cov}}

\def\bfnul{\mathbf 0}

\def\hbfx{\hat{\bfx}}
\def\hbfz{\hat{\bfz}}

\def\bbfx{\bar{\bfx}}

\def\hbfP{\hat{\bfP}}

\def\tbfx{\tilde{\bfx}}

\def\bfxi{\ensuremath{\boldsymbol{\xi}}}

\usepackage{hyperref}

\def\BibTeX{{\rm B\kern-.05em{\sc i\kern-.025em b}\kern-.08em
    T\kern-.1667em\lower.7ex\hbox{E}\kern-.125emX}}
\begin{document}

\title{Stone Soup: ADS-B-based Multi-Target Tracking with Stochastic Integration Filter\\

\thanks{O. Straka has been in part supported by the Ministry of Education, Youth and Sports of the Czech Republic under project ROBOPROX - Robotics and Advanced Industrial Production
CZ.02.01.01/00/22\_008/0004590. J. Duník and J. Matoušek have been in part supported by the Czech Science Foundation (GACR) under grant GA 25-16919J.}
}

% \author{\IEEEauthorblockN{John Hiles}
% \IEEEauthorblockA{\hspace*{1.5cm}\textit{Dept. of ECE}\hspace*{1.5cm} \\
% \textit{Virginia Commonwealth University}\\
% Richmond, VA, USA\\
% hilesj@vcu.edu}
% \and
% \IEEEauthorblockN{Jakub Matoušek}
% \IEEEauthorblockA{\hspace*{1.0cm}\textit{Dept. of Cybernetics}\hspace*{1.0cm} \\
% \textit{Univ. of West Bohemia}\\
% Pilsen, Czech Republic \\
% matoujak@kky.zcu.cz}
% \and
% \IEEEauthorblockN{Erik Blasch}
% \IEEEauthorblockA{\hspace*{2cm}\textcolor{red}{TBD}\textit{MOVEJ Analytics}\hspace*{2cm} \\
% Fairborn, OH USA \\
% erik.blasch@gmail.com}
% \and
% \IEEEauthorblockN{Ruixin Niu}
% \IEEEauthorblockA{\hspace*{1cm}\textit{Dept. of ECE} \hspace*{1cm}\\
% \hspace*{1cm}\textit{Virginia Commonwealth University\hspace*{1cm}}\\
% Richmond, VA, USA \\
% rniu@vcu.edu}
% \and
% \IEEEauthorblockN{Ondřej Straka}
% \IEEEauthorblockA{\hspace*{0.1cm}\textit{Dept. of Cybernetics}\hspace*{0.1cm} \\
% \textit{Univ. of West Bohemia}\\
% Pilsen, Czech Republic \\
% straka30@kky.zcu.cz}
% \and
% \IEEEauthorblockN{\hspace*{0cm}Jindřich Duník}
% \IEEEauthorblockA{\hspace*{2.2cm}\textit{Dept. of Cybernetics}\hspace*{2.2cm} \\
% \textit{\hspace*{0cm}Univ. of West Bohemia}\\
% \hspace*{0cm}Pilsen, Czech Republic \\
% \hspace*{0cm}dunikj@kky.zcu.cz}
% }

\author{
    \IEEEauthorblockN{John Hiles\IEEEauthorrefmark{1}, Jakub Matoušek\IEEEauthorrefmark{2}, Erik Blasch\IEEEauthorrefmark{3}, Ruixin Niu\IEEEauthorrefmark{1}, Ondřej Straka\IEEEauthorrefmark{2}, Jindřich Duník\IEEEauthorrefmark{2}} 
    \IEEEauthorblockA{\IEEEauthorrefmark{1}Dept. of ECE, Virginia Commonwealth Univ., Richmond, VA, USA \\ 
    \{hilesj, rniu\}@vcu.edu}
    \IEEEauthorblockA{\IEEEauthorrefmark{2}Dept. of Cybernetics, Univ. of West Bohemia, Pilsen, Czech Republic \\ 
    \{matoujak, straka30, dunikj\}@kky.zcu.cz}
    \IEEEauthorblockA{\IEEEauthorrefmark{3}MOVEJ Analytics, Fairborn, OH, USA \\ erik.blasch@gmail.com}
}

% \author{\IEEEauthorblockN{John Hiles}
% \IEEEauthorblockA{\hspace*{0cm}\textit{Dept. of ECE}\hspace*{0cm} \\
% \textit{Virginia Commonwealth Univ.}\\
% Richmond, VA, USA\\
% hilesj@vcu.edu}
% %\hspace*{5cm}
% \and
% \IEEEauthorblockN{Jakub Matoušek}
% \IEEEauthorblockA{\hspace*{0cm}\textit{Dept. of Cybernetics}\hspace*{0cm} \\
% \textit{Univ. of West Bohemia}\\
% Pilsen, Czech Republic \\
% matoujak@kky.zcu.cz}
% %\hspace*{5cm}
% \and
% \IEEEauthorblockN{Erik Blasch}
% \IEEEauthorblockA{\hspace*{0cm}\textit{MOVEJ Analytics}\hspace*{0cm} \\
% Fairborn, OH USA \\
% erik.blasch@gmail.com}
% %\hspace*{7cm}
% \and
% \IEEEauthorblockN{Ruixin Niu}
% \IEEEauthorblockA{\hspace*{0cm}\textit{Dept. of ECE} \hspace*{0cm}\\
% \hspace*{0cm}\textit{Virginia Commonwealth Univ.\hspace*{0cm}}\\
% Richmond, VA, USA \\
% rniu@vcu.edu}
% %\hspace*{3cm}
% \and
% \IEEEauthorblockN{Ondřej Straka}
% \IEEEauthorblockA{\hspace*{0cm}\textit{Dept. of Cybernetics}\hspace*{0cm} \\
% \textit{Univ. of West Bohemia}\\
% Pilsen, Czech Republic \\
% straka30@kky.zcu.cz}
% %\hspace*{2cm}
% \and
% \IEEEauthorblockN{\hspace*{0cm}Jindřich Duník}
% \IEEEauthorblockA{\hspace*{0cm}\textit{Dept. of Cybernetics}\hspace*{0cm} \\
% \textit{\hspace*{0cm}Univ. of West Bohemia}\\
% \hspace*{0cm}Pilsen, Czech Republic \\
% \hspace*{0cm}dunikj@kky.zcu.cz}
% }

\maketitle

\begin{abstract}
This paper focuses on the multi-target tracking using the Stone Soup framework. In particular, we aim at evaluation of two multi-target tracking scenarios based on the simulated class-B dataset and ADS-B class-A dataset provided by OpenSky Network. The scenarios are evaluated w.r.t. selection of a local state estimator using  a range of the Stone Soup metrics. Source code with scenario definitions and Stone Soup set-up are provided along with the paper.
\end{abstract}


\section{Introduction}
In the past few decades, there has been a lot of interest in nonlinear state estimation of stochastic dynamic systems using noisy observations. Nonlinear state estimators can be largely divided into two categories: global and local filters. Global filters, such as the ensemble Kalman filter (EnKF) \cite{Eve:09}, point-mass filter \cite{DuSoVeStHa:19}, and particle filter \cite{RiAruGo:04}, provide the approximated posterior  probability density function (PDF) of the system state. On the other hand, to take advantage of the Kalman filter's recursive framework and achieve lower computational complexity, local filters, such as the extended Kalman filter (EKF) \cite{bar-shalom&etal:book11}, unscented Kalman filter (UKF) \cite{JuUhlDu:00}, cubature Kalman filter (CKF) \cite{ArHa:09}, and Monte Carlo Kalman filter (MCKF) \cite{So:00}, calculate the conditional mean and covariance matrix instead of the posterior PDF.   

In comparison to other local filters, the \textit{stochastic integration filter} (SIF) \cite{DuStrSi:13}, offers an asymptotically accurate integral evaluation of the conditional mean and covariance matrix, with reasonable computation complexity. When selecting some of the SIF's random parameters deterministically, the UKF and CKF are special cases of the SIF. The SIF has been shown to provide improved tradeoff between estimation accuracy and computational complexity, in comparison to the EnKF, MCKF, EKF, and UKF \cite{DuStSiBl:15,Dunik&etal:fusion_24}.  In \cite{Dunik&etal:fusion_24}, the SIF was implemented in Python and incorporated in the Stone Soup framework. However,  the testing  scenarios in \cite{Dunik&etal:fusion_24} only involve a single target and  simulated data.  The focus of this paper is to investigate the performance of the SIF for multi-target tracking (MTT) using both simulated and real automatic dependent surveillance–broadcast (ADS-B) data in the Stone Soup framework. 

\subsection{Multitarget Tracking in ABS-B}

%Demonstration of the stochastic integration filter in multi-target tracking.

MTT has widely been explored since the 1960’s to support surveillance, navigation, and control. Popular MTT algorithms include the joint probabilistic data association filter (JPDAF) \cite{bar-shalom&etal:book11}, multiple hypothesis tracking (MHT) \cite{blackman&popoli:book99}, random finite set (RFS) based filters \cite{mahler:book07,mahler:book14,Vo&etal:ieee_sp14}, and message passing based algorithms \cite{meyer&etal:ieee_proceedings18}. Typically, MTT includes measurement pre-processing, estimation, and prediction to monitor a coverage area. Measurement pre-processing is usually applied to increase the signal-to-noise ratio (SNR).  Estimation includes the refinement of the state to reduce the error and prediction uses the current state estimate to predict the future expected measurement location. Among the MTT steps, resolving the measurement to the correct target is data association which is a key challenge.  Hence, the associations within MTT include measurement-to-measurement association, measurement-to-track association, and track-to-track association. To research these needs among filtering techniques, scenarios, and metrics, the ISIF community has developed the open source Stone Soup collection. Managed by the ISIF Open Source Tracking and Estimation Working Group (OSTEWG)\cite{ostewg2025stonesoup}, an emerging set of scenarios and metrics afford the comparison of contemporary methods – such as the SIF. %The goal was to focus on the SIF capability and development, whereas using the Stone Soup simulator, the data association was reasonably assumed to reliable for track-to-truth.

For this paper, the scenario of choice was to research the ADS-B scenario based on the OpenSky data. ADS-B data has grown in popularity for research since its mandated use for aerospace systems in the early 2010s for full scale operation in 2020.  Currently, the data is accessible from FightRadar24 to track all the aircraft in the sky that are reporting.  Hence, with the ADS-B, methods for air traffic management (ATM) can be assessed to improve efficiency, safety, and reliability for increasing the density of airspace that includes military (e.g., surveillance), commercial (e.g., passengers), and business operations (e.g., air package delivery). At the same time, since the ADS-B is air traffic, similar constructs for ground (transponders) and maritime (automatic identification system  AIS) are being considering for space operations for space traffic management (STM). With such a capability, research is emerging on the efficacy of the methods in response to measurement attacks of spoofing, jamming, and replay.

% As related to OpenSky \cite{SchStLe:14};, highlight some limitations encountered when collecting the ADS-B data such as (1) antenna positions, (2) bandwidth variations, (3) local policies, (4) data repository, and (5) interval rates. Hence, research challenges include coverage, GNSS validation, and decoding. As with the attacks of the ADS-B from nefarious activities, navigation, control, and surveillance could be altered necessitating a need for MTT methods to ensure the integrity of the results.

\subsection{Goal of the Paper}
In this paper, we simulate two scenarios using the Stone Soup framework;
\begin{itemize}
    \item[\textit{(i)}] Simulation involving several overlapping trajectories. In this scenario, the bearing, range, and elevation measurements are sequential and without clutter.
    \item[\textit{(ii)}] Usage of the OpenSky aircraft ADS-B data as ground truth and simulation of the bearing, range, elevation detections from multiple sensors introducing the clutter.
\end{itemize}
These two MTT scenarios are implemented using the EKF, UKF, and the SIF selected as the state estimators. The MTT performance is evaluated using multiple metrics including the optimal sub-pattern assignment and single integrated air picture position accuracy. The source codes are provided along with the paper\footnote{https://github.com/0sm1um/ADS-B-Tracking}.

The rest of the paper is organized as follows. In Section~II the Stone Soup framework is reviewed. Section III focuses on brief introduction of the MTT with an emphasis on the local filter design. Then, Section IV describes considered blocks of Stone Soup used in the numerical experiments. The experiments are discussed in Sections V and VI for class-B aerospace  and ADS-B datasets, respectively. The concluding remarks are drawn in Section VII.

\section{Stone Soup}
\begin{figure}
    \centering
    \includegraphics[width=0.95\linewidth]{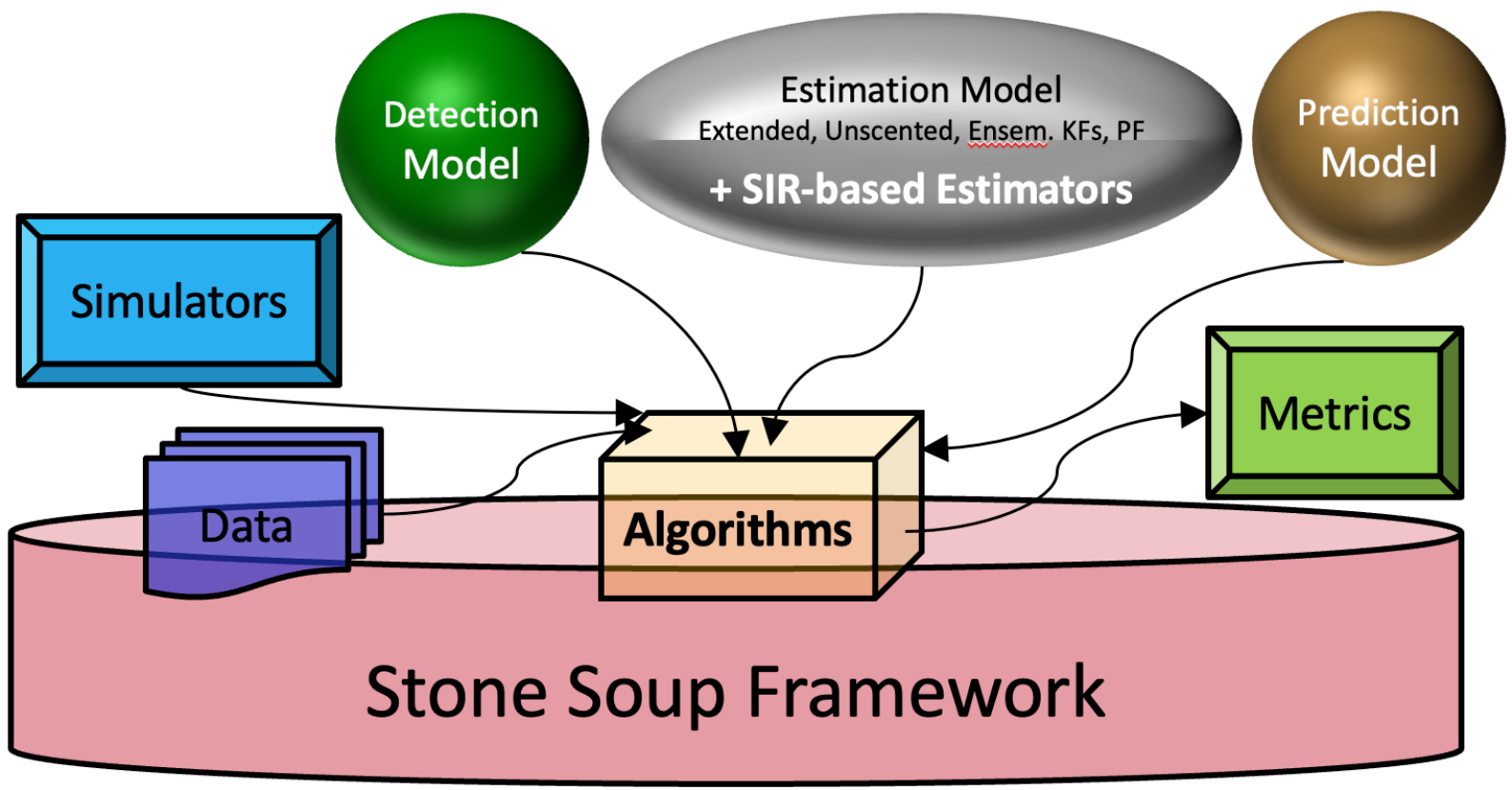}
    \caption{Elements of the ISIF Stone Soup framework \cite{Dunik&etal:fusion_24}.}
    \label{fig:stonesoup}
    \vspace*{-0mm}
\end{figure}

The Stone Soup project project is an open-source tracking and estimation framework currently available as a python library; to support developers, users, and systems engineers. The framework saw its first public alpha release in 2017 \cite{ThomasPaulA2017Aosf} with the first beta release in 2019. The name of the framework is inspired by the tale of the “Stone Soup”; from European folklore in which many ingredients (estimation methods) are contributed by villagers (researchers) to devise a flavourful soup(useful software). Inherent in the common repository is the ability to compare and reuse contributions to extend the quality of the estimation capabilities. While some have sought initial ideas \cite{BlaschEruj2020prc, ORourkeSeanM2021IoEK}, Barr et al. \cite{BarrJordi2022SSos} highlight the many capabilities such as track generation, filtering, classification, data association, and sensor management routines. Since then the Stone Soup has been extended with a variety of components and routines \cite{CoPrLiGaKiGo:24,Dunik&etal:fusion_24,WrSuDaPrHo:24} and offers the following tools for end-to-end solution of the MTT task: %Examples are provided for video tracking, automatic dependent surveillance – broadcast (ADS-B) tracking, orbital estimation, drone tracking, sensor management, and tracking evaluation. Recently, a series of papers began using the techniques, specifically for radar-based tracking. Carniglia et al. \cite{BalajiBhashyam2022IoSB} used the Stone Soup Generalized Optimal Sub-Pattern Assignment (GOSPA) and Single Integrated Air Picture (SIAP) metrics to compare the Joint-Probabilistic Data Association (JPDA) to track multiple airborne targets being in clutter from two ground-based radars to assess the track quality sensitivity of biased sensor measurements. Similarly, tracking of the adaptive Kernel Kalman Filter (AKKF) was compared to the particle filter (PF) using Stone Soup infrastructure \cite{WrightJames0256739}. In 
%In the current version the Stone Soup offers a variety of routines and tools such as
\begin{enumerate}
  \item \textbf{Track generator} - For comparative analysis, the truth generator enables track initiation, trajectory analysis, and modifications such as emulating an outage.
  \item \textbf{Track association} - A simulator provides control over the data association between measurements and targets.
  \item \textbf{Track metrics} - The Stone Soup repository offers a variety of contemporary measures, including:
  \begin{itemize}
    \item Weighted Euclidean distance measure,
    \item Squared Mahalanobis distance measure,
    \item Squared Gaussian Hellinger distance measure,
    \item Kullback–Leibler divergence,
    \item Generalized Optimal Subpattern Assignment.
  \end{itemize}
  The Standardized Information and Analysis Program measures includes:
  \begin{itemize}
    \item Ambiguity,
    \item Completeness,
    \item Longest tracking,
    \item Track rate spuriousness,
    \item Velocity and position accuracy.
  \end{itemize}
  \item \textbf{Track estimators} - Considering measurement-to-track association, the target state can be estimated by a number of filters offering a trade-off between complexity and accuracy, such as:
  \begin{itemize}
    \item EKF, UKF, SIF,
    \item Particle filters.
  \end{itemize}
  Additional features include track deletion, handling out-of-order sequences, and density management.
\end{enumerate}
The overall scheme of the Stone Soup framework with the SIR-based estimators is illustrated in Fig. \ref{fig:stonesoup}.

%The simulation was conducted under various scenarios to assess the effects of omitting track generation, track initiation, and measurement-to-track association. Using the ADS-B scenario from Stone Soup, the parameters and selections were based on that analysis.

\section{Multiple Target Tracking and State Estimation}
Multitarget tracking is a well-developed and wide area that aims at estimation of the state an unknown number $n_k\geq 0$ of targets based on the a number $m_k\geq 0$ of measurements at time-step $k$. Whereas the state includes for example, position and velocity, of each tracked target (or object) and is unknown, the measurement can be seen as an available noisy nonlinear transformation of the state, for example in the form of range and bearing of each target w.r.t. a radar. As the measurements are imperfect and cluttered, the number of tracked targets $n_k$ and available measurements $m_k$ need not be equal and has to be associated. This measurement to target state association can be addressed by a variety of methods including the well-known and in the Stone Soup implemented JPDAF and MHT algorithms \cite{BarrJordi2022SSos}. %probabilistic data association (PDA) filter for tracking a single target in clutter, its extension for a variable number of targets known as joint PDA (JPDA) filter, or a  GNN maximizing the likelihood of association using a squared Mahalanobis distance [REF TBD]. \textcolor{red}{Unify abbreviations with introduction - to check :-)} 

Having the association performed, that the state of each target  is estimated by a dedicated, in our case local, filter. In this paper, we focus on the assessment of the local filter selection on the performance of the whole MTT system. Therefore, in the following parts, we briefly introduce the state-space model for the target modeling and the local state estimator design for the its state estimation.

\subsection{State-Space Model}
The target motion can be conveniently described by a discrete-time stochastic dynamic state-space model with linear dynamics and nonlinear measurement equation
\begin{align}%\hspace*{10mm}
    \bfx_{k+1}&=\bfF_{k}\bfx_{k}+\bfw_{k}, \label{eq:asx}\\
	\bfz_{k}&=\bfh_k(\bfx_k)+\bfv_k, \label{eq:asz}
\end{align}
where the vectors $\bfx_k\in\real^{n_x}$ and $\bfz_k\in\real^{n_z}$ represent the \textit{unknown and sought} state of the system and the \textit{available} measurement at time instant $k$, respectively, and $k=0,1,2,\ldots,T$.  The state transition matrix $\bfF_k:\real^{n_x\times n_x}$ and the measurement function $\bfh_k:\real^{n_x}\rightarrow\real^{n_z}$ are supposed to be known. The state dynamics can be described by a number of models including nearly constant velocity, known/unknown turn-rate, or a Singer models, to name a few. The measurement equation depends on the sensors employed. If a radar is used, then the measurement consists of the range and bearing angle between the target and the radar.

The state noise $\bfw_k\in\real^{n_x}$, the measurement noise $\bfv_k\in\real^{n_z}$, and the initial state $\bfx_0\in\real^{n_x}$ are supposed to be independent of each other. The noises and the initial state are assumed to be normally distributed, i.e.,
\begin{align}
	p(\bfw_k)&=\calN\{\bfw_k;\bfnul_{n_x\times1},\bfQ_k\},\label{eq:pw}\\
	p(\bfv_k)&=\calN\{\bfv_k;\bfnul_{n_z\times1},\bfR_k\},\label{eq:pv}\\
	p(\bfx_0)&=\calN\{\bfx_0;\bbfx_0,\bfP_0\},\label{eq:px0}
\end{align}
where $\bfnul_{n_x\times1}\in\real^{n_x\times1}$ is a zero vector and the notation $\calN\{\bfx;\bbfx,\bfP\}$ stands for the Gaussian PDF of a random variable $\bfx$ with mean $\bbfx$ and covariance matrix $\bfP$. The first two moments of the random variables in \eqref{eq:pw}--\eqref{eq:px0} are supposed to be known.

% \subsection{Problem Formulation}\label{AA}
% Sequential and Synchronous Measurements

% Consider the discrete time stochastic system
% \begin{equation}
% \mathbf{x}_{k+1} = f(\mathbf{x}_k) + \mathbf{\nu}_k\label{eq:state}
% \end{equation}

% With measurement equation
% \begin{equation}
% \mathbf{z}_{k+1} = h(\mathbf{x}_k) + \mathbf{\xi}_k\label{eq:meas}
% \end{equation}

% With normally distributed noise
% \begin{align}
% \mathbf{\nu}_k \sim \mathcal{N}(0, Q) && \mathbf{\xi}_k \sim \mathcal{N}(0, R) \label{eq:noise}
% \end{align}

% \subsubsection{Track Initiation}
% Global Nearest Neighbors, Mahalanobis Distance

% Mahalanobis Distance

% \begin{align}
%     d^2 = [z - \hat{z}_{k+1|k}]^T S_{k+1}^{-1} [z-\hat{z}_{k+1|k}]
% \end{align}

% \subsubsection{Track Deletion}
% Tracks are deleted if the trace of the Covariance matrix exceeds a certain threshold. This is an important factor because it highlights that the consistancy of the SIF will allow it to maintain tracks where the EKF or Particle Filter would lose them.

\subsection{Extended, Unscented, Cubature, and Stochastic Integration Filters}
The EKF, UKF, CKF, and SIF are representatives of the \textit{local approaches} to state estimation \cite{Sa:13,SiDu:09}. In particular, the local filters provide the estimate of the state $\bfx_k$ conditioned on all available measurements $\bfz^k=[\bfz_0,\bfz_1,\ldots,\bfz_k]$ and the state-space model \eqref{eq:asx}--\eqref{eq:px0} in the form of the filtering conditional mean and covariance matrix defined as
\begin{align}
    \hbfx_{k|k}=\mean[\bfx_k|\bfz^k],\\
    \hbfP^{xx}_{k|k}=\cov[\bfx_k|\bfz^k].
\end{align}
In the Bayesian framework, the mean and the covariance matrix can be interpreted as a moments defining the \textit{approximate} \textit{Gaussian} filtering PDF 
\begin{align}
    p(\bfx_k|\bfz^k)\approx\calN\{\bfx_k;\hbfx_{k|k},\hbfP^{xx}_{k|k}\}.
\end{align}

Respecting the Bayesian perspective for local filter design, the algorithm of any local filter reads:

\noindent\rule[-\baselineskip]{\linewidth}{1pt}
\vspace{-4pt}

\textbf{Algorithm:} Generic Local Filter\\ %\textcolor{red}{TBD check consistency of notation}\\
\noindent\rule[\baselineskip]{\linewidth}{1pt}
\vspace{-1cm}
\begin{descriptionA}
	\item Set the time instant $k=0$ and define an 
	initial condition $p(\bfx_0|\bfz^{0})=\calN\{\bfx_0;\hbfx_{0|-1}, \bfP^{xx}_{0|-1}\}$.	 
	\item The moments of the filtering estimate $p(\bfx_{k}|\bfz^{k})\approx\calN\{\bfx_{k};\hbfx_{k|k}, \bfP^{xx}_{k|k}\}$ are %given by
	\begin{align}
		\hbfx_{k|k}&=\hbfx_{k|k-1}+\bfK_{k}(\bfz_{k}-\hbfz_{k|k-1})\enspace,\label{eq:LFxfB}\\
		\bfP^{xx}_{k|k}&=\bfP^{xx}_{k|k-1}-\bfK_{k}\bfP^{zz}_{k|k-1}\bfK_{k}^T\enspace,\label{eq:LFPfB}
	\end{align}
	where $\bfK_{k}=\bfP^{xz}_{k|k-1}(\bfP^{zz}_{k|k-1})^{-1}$ is the   
	filter gain,   
	\begin{align}
		\hbfz_{k|k-1}&\!\!=\!\!\int\!\!\bfh_{k}(\bfx_{k})\calN\{\bfx_{k};\hbfx_{k|k-1}, \bfP^{xx}_{k|k-1}\}d\bfx_{k},\label{eq:LFzpB}\\     
		\bfP^{zz}_{k|k-1}&=\int(\bfh_{k}(\bfx_{k})-\hbfz_{k|k-1})(\cdot)^T\nonumber\\
		&\times\calN\{\bfx_{k};\hbfx_{k|k-1}, \bfP^{xx}_{k|k-1}\}d\bfx_{k}+\bfR_{k}, \label{eq:LFPzB}\\
		\bfP^{xz}_{k|k-1}&=\int(\bfx_{k}-\hbfx_{k|k-1})(\bfh_{k}(\bfx_{k})-\hbfz_{k|k-1})^T\nonumber\\
		&\times\calN\{\bfx_{k};\hbfx_{k|k-1}, \bfP^{xx}_{k|k-1}\}d\bfx_{k}. \label{eq:LFPxzB}
	\end{align}
    \item The predictive moments of the Gaussian-assumed PDF $p(\bfx_{k+1}|\bfz^{k})\approx\calN\{\bfx_{k+1};\hbfx_{k+1|k}, \bfP^{xx}_{k+1|k}\}$ are calculated according
    \begin{align}
        \hbfx_{k+1|k}&=\mean[\bfx_{k+1}|\bfz_k]=\bfF_k\hbfx_{k|k},\label{eq:LFxpB}\\     
		\bfP^{xx}_{k+1|k}&=\mean[\bfx_{k+1}|\bfz_k]=\bfF_k\bfP_{k|k}\bfF_k^T+\bfQ_k.\label{eq:LFPxB}
  %       \hbfx_{k+1|k}&\!\!=\!\!\int\!\!\bff_{k}(\bfx_{k})\calN\{\bfx_{k};\hbfx_{k|k}, \bfP^{xx}_{k|k}\}d\bfx_{k},\label{eq:LFxpB}\\     
		% \bfP^{xx}_{k+1|k}&=\int(\bff_{k}(\bfx_{k})-\hbfx_{k+1|k})(\cdot)^T\nonumber\\
		% &\times\calN\{\bfx_{k};\hbfx_{k|k}, \bfP^{xx}_{k|k}\}d\bfx_{k}+\bfQ_{k}.\label{eq:LFPxB}
    \end{align}
    
    The algorithm then continues to {\bf Step 2} with $k\hookleftarrow k+1$.
\end{descriptionA}
\noindent\rule[\baselineskip]{\linewidth}{1pt}

The evaluation of the expected values \eqref{eq:LFzpB}--\eqref{eq:LFPxzB} for the measurement prediction in the filtering step 
% \begin{itemize}
%     % \item \eqref{eq:LFxpB}, \eqref{eq:LFPxB} for prediction, and 
%     \item \eqref{eq:LFzpB}--\eqref{eq:LFPxzB} for the measurement prediction in the filtering step,
% \end{itemize}
can be seen as a calculation of the moments of a nonlinearly transformed Gaussian random variable with known description \cite{SiDu:09,NoPoRa:00b}. This means that calculation of the moments can be interpreted as an evaluation of a \textit{Gaussian-weighted} integral
\begin{align}
    \mathcal{I}=\mean[\bfg(\bfx_k)|\bfz^{k-1}]=\int \bfg(\bfx_k) \calN\{\bfx_k;\hbfx_{k|k-1},\bfP_{k|k-1}^{xx}\}d\bfx_k,\label{eq:int}
\end{align}
where the conditional mean $\hbfx_{k|k-1}$ and the covariance matrix $\bfP_{k|k-1}^{xx}$ are known from estimator predictive steps and the nonlinear function is \textit{(i)} $\bfg(\bfx_{k})=\bfh_k(\bfx_{k})$ for measurement prediction $\hbfz_{k|k-1}$ \eqref{eq:LFzpB} calculation, \textit{(ii)} $\bfg(\bfx_{k})=(\bfh_{k}(\bfx_{k})-\hbfz_{k|k-1})(\bfh_{k}(\bfx_{k})-\hbfz_{k|k-1})^T$ for measurement predictive covariance matrix $\bfP^{zz}_{k|k-1}$ \eqref{eq:LFPzB} calculation, and \textit{(iii)} $\bfg(\bfx_{k})=(\bfx_{k}-\hbfx_{k|k-1})(\bfh_{k}(\bfx_{k})-\hbfz_{k|k-1})^T$ for state and measurement predictive covariance matrix  $\bfP^{xz}_{k|k-1}$ \eqref{eq:LFPxzB} calculation.

%stems from the model and calculated moment definition (namely mean or covariance matrix evaluation). %The index $\ell$ stands for $k$ if the state prediction \eqref{eq:LFxpB}, \eqref{eq:LFPxB} is calculated and for $k-1$ if the measurement prediction \eqref{eq:LFzpB}--\eqref{eq:LFPxzB} is calculated.

Closed-form solution to the Gaussian weighted integrals is possible for a narrow class of functions $\bfh_k(\cdot)$, such as the \textit{linear} function for which any local filter becomes the optimal Kalman filter. For nonlinear case, an approximate solution has to be used. In the area of the local filters two principal approximations can be used, namely \textit{(i)} linearisation of nonlinear function $\bfg(\cdot)$ allowing an \textit{analytical} solution to \eqref{eq:int}, \textit{(ii)} approximation of the conditional Gaussian-assumed PDF $\calN\{\bfx_k;\hbfx_{k|k-1},\bfP_{k|k-1}^{xx}\}$ by a set weighted points enabling a \textit{numerical} solution to \eqref{eq:int}.

The \textit{former} solution is based on the linearisation of the nonlinear function $\bfg(\cdot)$ by the Taylor expansion or the Stirling's interpolation, which leads e.g., to the EKF, second order filter, or the divided difference filters \cite{AnMo:79,Sa:13,NoPoRa:00}. The \textit{latter} solution to \eqref{eq:int} applies a quadrature, cubature, or stochastic numerical integration rule of the form 
\begin{align}
\mathcal{I}\approx\hat{\mathcal{I}}=\sum_{i=1}^S\omega^{(i)}\bfg(\bfxi_{k|k-1}^{(i)}),\label{eq:sir}
\end{align}
where $\bfxi_{k|k-1}^{(i)}$ and $\omega^{(i)}$ are suitably chosen spherical-radial points and corresponding weights of the rule, respectively. This class of local filters is represented by the SIF, unscented or cubature Kalman filters \cite{JuUhl:04,ArHa:09,DuSiStr:12,DuStSiBl:15}. Although both approximate solutions are based on different ideas and approximations, the resulting algorithms can be written in a unified form \cite{SiDu:09}. The computational complexity of all local filters is $\calO(n_x^3+n^3_z)$, but it depends on the particular implementation \cite{GreAnd:01}.

%generally the filters based on Stirling's interpolation and numerical integration rules are more demanding than those based on the Taylor expansion due to necessity to calculate the factor of the covariance matrix at each time step. Detailed analysis of local filter computational complexity can be %found in \cite{GreAnd:01}.

\subsection{Expected Performance of Filters and the Goal of the Paper}
The EKF is probably the most used algorithm in the tracking and navigation, and this is due to several reasons; it is \textit{(i)} a widely known algorithm (for a couple of decades), \textit{(ii)} computationally efficient and with a plethora of stable implementations \cite{AnMo:79,Si:06}, and \textit{(iii)} a certifiable algorithm for a civil aircraft navigation system \cite{DO229D:06}. On the other hand, the EKF performance may degrade for models with a higher degree of nonlinearity or higher uncertainty, where the linearisation is not valid anymore. In this case, the EKF tends to underestimate the estimate covariance matrix $\bfP_{k|k}$ (in comparison to the optimal or "true" Nonlinear Filter) which leads to an overly optimistic impression of accuracy which does not correspond to the true mean estimate error $\tbfx_{k|k}=\bfx_k-\hbfx_{k|k}$. Based on extensive analysis and numerical verification, the SIF seems to be a more robust algorithm (not only w.r.t. EKF, but also UKF and CKF) for models with a higher degree of nonlinearity providing a consistent estimate. \cite{DuStSiBl:15,ZhGu:17,Dunik&etal:fusion_24}. The reason can be found in a more realistic assessment of the estimate uncertainty. On the other hand, the SIF has higher computational complexity.

The \textit{goal} of the paper is the assess the impact of the selected local filters on the performance of the MTT system which might be seen as a complex nonlinear problem. The assessment is performed using the Stone Soup framework with simulated and OpenSky Network downloaded data.

\section{Stone Soup Implementation Description}
Having the filters introduced, we briefly describe the overall setting of the multi-target tracking task in the Stone Soup framework. Description of the Stone Soup components used in our implementation and simulations, including the tracker architecture and the metrics employed for performance evaluation, follows.

\subsection{Tracker}
The implemented MTT system employs the global nearest neighbour (GNN) data association method with a 2D assignment algorithm. This setup ensures efficient and accurate association between predicted tracks and incoming detections. The classes used for the MTT tracker are:
\begin{itemize}
    \item \textbf{Data Associator} (routine \textit{GNNWith2DAssignment}):  
    Utilizes the GNN to associate detections (or radar measurements) to predicted target states. This method constructs a 2D matrix of distances between predictions and detections and solves the assignment problem to find the optimal associations.
    \item \textbf{Distance Hypothesiser}: Generates assignement hypotheses based on the distance between predicted and detected states, using the Mahalanobis distance measure. \textit{Parameters}:
    \begin{itemize}
        \item Mahalanobis distance calculates the distance between predicted and detected states.
        \item Missed detection distance threshold is set to 5. If the distance between a predicted state and a detection exceeds this value, the detection is considered a missed detection. This parameter helps in managing associations in cluttered environments by limiting the consideration of unlikely associations.
    \end{itemize}
    \item \textbf{Target Deleter}: Deleter is a component used to remove targets from the tracking system based on the time since their last prediction or update. It is configured with a time threshold, and targets are deleted if they have not been updated or predicted within this threshold period. \textit{Parameters}:
    \begin{itemize}
        \item Time threshold is set to $t=10$.
    \end{itemize}
\end{itemize}

This configuration ensures robust tracking performance by effectively managing data association, even in the presence of clutter and missed detections. Although for the purpose of this paper, we are not considering clutter.

\subsection{Filters}
The MTT system is designed with two filters, namely the \textit{standard} EKF and the \textit{robust} third-order SIF \cite{Dunik&etal:fusion_24}. Note that the UKF, when run with the default parameters in Stone Soup, experienced stability issues, resulting in a covariance matrix that was not semidefinite positive.  %Note that the UKF and the CKF provide very similar performance to the EKF in the selected scenarios.%perform similar with the EKF in the considered %\textcolor{red}{note on UKF failing?}

\subsection{Metrics}
To evaluate the performance of the MTT system in the presence of clutter, we employ the following standard metrics\footnote{\url{https://stonesoup.readthedocs.io/en/latest/auto_examples/metrics/Metrics.html}} from the Stone Soup framework:
\begin{itemize}
    \item \textbf{Optimal Sub-Pattern Assignment (OSPA)}:  
    The OSPA metric quantifies the accuracy of the estimated target states compared to the ground truth by considering both localization errors and cardinality errors. It provides a balanced assessment of multi-target tracking performance, penalizing both position estimation errors and missed or false detections.    
    \textit{Parameters}: 
    \begin{itemize}
        \item  Norm associated to distance is $p$.
         \item Maximum distance for possible association is $c$.
    \end{itemize}
    Parameter $p$ can be generally thought of as the sensitivity to outliers\cite{RaFeSv:16} in tracks. $p=2$ is used for both examples. Parameter $c$ is associated with the degree to which Cardinality errors dominate the metric. Lower values of $c$ tend to favor the precision of tracks, and higher values disregard locality errors. Optimal choice of $c$ primarily depends on the magnitude of a typical locality error (distances between a track and a truth) and hence the Class-B Airspace example uses $c=10$ and the Class-A Airspace example uses $c=250$.

    \item \textbf{Single Integrated Air Picture (SIAP) Ambiguity} \cite{Vo:01}: 
    The SIAP ambiguity metric assesses the number of tracks assigned to a true object. The score is defined as
    
    \begin{align}
        A = \quad \frac{\sum_{k=0}^{T}N_{A,k}}{\sum_{k=0}^{T}J_{T,k}},
    \end{align}
    where $N_{A,k}$ represents the number of associated tracks at a given time, and $J_{T,k}$ represents the number of associated true targets. A score of 1 is optimal.
    
    \item \textbf{Single Integrated Air Picture (SIAP) Position Accuracy} \cite{Vo:01}: 
    The SIAP position accuracy metric assesses the mean positional error of the estimated tracks relative to ground truth. The Positional Accuracy Metric is given as:
    \begin{align}
        PA = \frac{\sum_{k=0}^{T}{\sum_{n\in D_k}PA_{n,k}}}
                        {\sum_{k=0}^{T}{N_{A,k}}},
    \end{align}
    where $D_k$ is the set of tracks held at timestamp $k$ and $PA_{n,k}$ is the Euclidean distance of track n to its associated truth at the corresponding timestamp.
    
    \item \textbf{Sum of Covariance Norms}:
    This metric evaluates the overall uncertainty in state estimation by summing the Frobenius norms of the estimated covariance matrices across all tracked targets. A lower value indicates more confident and precise state estimates, while a higher value suggests greater uncertainty in tracking results \cite{RandallT.J.1985McbG}. The Frobenius Norm used and corresponding metric are given as:
    \begin{align*}
        ||\bfP_{k|k}^{xx}||_{F} = \sqrt{\trace((\bfP_{k|k}^{xx})^T \bfP_{k|k}^{xx})}.
    \end{align*}

\end{itemize}

These metrics collectively provide a comprehensive evaluation of the tracking performance, balancing accuracy, robustness, and uncertainty quantification in a cluttered MTT environment.

\section{Simulated Class-B Airspace Dataset}

Presented here is a scenario of crowded airspace, typically from the surface to $10,000\ [ft]$ above sea level. Class-B Airspace is a designation typically reserved for the busiest airports in a country. All aircraft operating in Class B Airspace typically need clearance from Air Traffic Controllers to fly, and Class-B Airspace is typically very closely monitored as it is the zone in which collisions are most probable.

\subsection{Simulation Parameters}
In this scenario, 10 maneuvering targets in a relatively enclosed 2D airspace are simulated. All targets start with randomly generated initial position within a $30\ [km]$ by $300\ [km^2]$. All targets have velocities in two dimensions randomly initialized between $\pm 200\ [m/s]$.

Each target state is defined as its position in $[m]$ and velocity in $[m/s]$ in north direction and in east direction, i.e., $\bfx_k=[p_{\mathrm{N},k}, v_{\mathrm{N},k}, p_{\mathrm{E},k}, v_{\mathrm{E},k}]^T=[\bfx_{1,k},\bfx_{2,k},\bfx_{3,k},\bfx_{4,k}]^T$, $\bfx_{i,k}$ stands for $i$-th element of the vector $\bfx_k$. The state of each target can follow one of the following \textit{three} motion models, namely
\begin{itemize}
    \item Nearly constant velocity (NCV) model with matrices defining the state equation \eqref{eq:asx}
    \begin{align}
    \bfF_k = \begin{bmatrix}
        1 & d t & 0 & 0\\
        0 & 1 & 0 & 0 \\
        0 & 0 & 1 & dt \\
        0 & 0 & 0 & 1
    \end{bmatrix}, %\bfQ_k = \begin{bmatrix}
    %     \tfrac{d t^3}{3} & \tfrac{d t^2}{2} & 0 & 0 \\
    %     \tfrac{d t^2}{2} & \tfrac{d t^3}{3} & 0 & 0 \\
    %     0 & 0 & \tfrac{dt^3}{3} & \tfrac{d t^2}{2} \\
    %     0 & 0 & \tfrac{d t^2}{2} & \tfrac{d t^3}{3}
    % \end{bmatrix}
    \bfQ_k = \begin{bmatrix}
        q_x\Sigma & \bfnul_2\\
        \bfnul_2 & q_y\Sigma
    \end{bmatrix},
    \end{align}
    where $\bfnul_2$ is zero matrix of indicated dimension, $\Sigma=\begin{bmatrix}
        \tfrac{d t^3}{3} & \tfrac{d t^2}{2} \\
        \tfrac{d t^2}{2} & dt
    \end{bmatrix}$, $dt=1\ [s] $ is a sampling period, and $q_x=q_y=0.05$.
    \item Known turn rate (TR) model with matrices 
    \begin{align}
    \bfF_k &= \begin{bmatrix}
              1 & \frac{\sin\omega dt}{\omega} &
              0 &-\frac{1-\cos\omega dt}{\omega} \\
              0 & \cos\omega dt & 0 & -\sin\omega dt \\
              0 & \frac{1-\cos\omega dt}{\omega} &
              1 & \frac{\sin\omega dt}{\omega}\\
              0 & \sin\omega dt & 0 & \cos\omega dt
          \end{bmatrix},\\%\bfQ_k &=  \begin{bmatrix}
          %     q_x \frac{dt^3}{3} & q_x \frac{dt^2}{2} &
          %     0 & 0 \\
          %     q_x \frac{dt^2}{2} & q_x dt &
          %     0 & 0 \\
          %     0 & 0 &
          %     q_y \frac{dt^3}{3} & q_y \frac{dt^2}{2}\\
          %     0 & 0 &
          %     q_y \frac{dt^2}{2} & q_y dt
          % \end{bmatrix}
          \bfQ_k &= \begin{bmatrix}
        q_x\Sigma & \bfnul_2\\
        \bfnul_2 & q_y\Sigma
    \end{bmatrix},
    \end{align}
    with the turn rate $\omega=\SI{20}{\degree}$.
    \item Known TR model with $\omega=-\SI{20}{\degree}$.
\end{itemize}
At any given time, targets have a finite probability of switching from one transition model to another according to transition probability matrix given by
\begin{align}
     \bfT = %\begin{bmatrix}
    %     t_{1,1} & t_{1,2} & t_{1,3} \\
    %     t_{2,1} & t_{2,2} & t_{2,3} \\
    %     t_{3,1} & t_{3,2} & t_{3,3} 
    % \end{bmatrix} =  
    \begin{bmatrix}
        0.7 & 0.15 & 0.15 \\
        0.4 & 0.6 & 0 \\
        0.6 & 0.4 & 0 
    \end{bmatrix}.
\end{align}

The target position is observed by the radar providing measurement of the range in $[m]$ and bearing in $[rad]$, which are related to the state using \eqref{eq:asz}, where
\begin{align}
    \bfh_k(\bfx_k)=\left[\begin{matrix}
        \sqrt{(\bfx_{1,k}-r_\mathrm{N})^2+(\bfx_{3,k}-r_\mathrm{E})^2} \\
        \arctan\tfrac{\bfx_{3,k}^2-r_\mathrm{E}}{\bfx_{1,k}^2-r_\mathrm{N}}
    \end{matrix}\right],
\end{align}
and $[r_\mathrm{N},r_\mathrm{E}]^T = [0,0]^T$ is the known radar position. The measurement is affected by the measurement noise $\bfv_k$ with the covariance matrix
\begin{align}
    \bfR=\left[\begin{matrix}4 & 0 \\ 0 & 0.5(\pi/180)^2\end{matrix}\right].
\end{align}

Detailed settings for the complete MTT system can be found linked github repository in the form of a Jupyter notebook.%\footnote{https://github.com/0sm1um/ADS-B-Tracking}

\subsection{Results}

\begin{figure}[]
	\centering
	\includegraphics[width=1\linewidth]{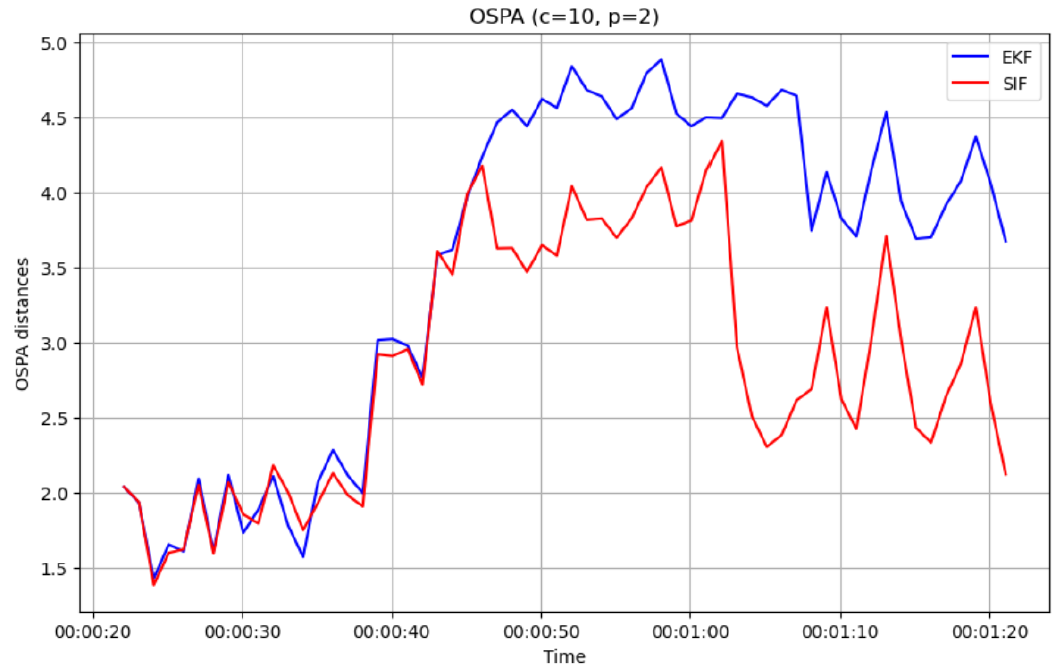}\vspace*{-2mm}
	\caption{OSPA metrics results for Class-B example.}
	\label{fig:OSPA1}\vspace*{-4mm}
\end{figure} 

\begin{figure}[] 
	\centering
	\includegraphics[width=1\linewidth]{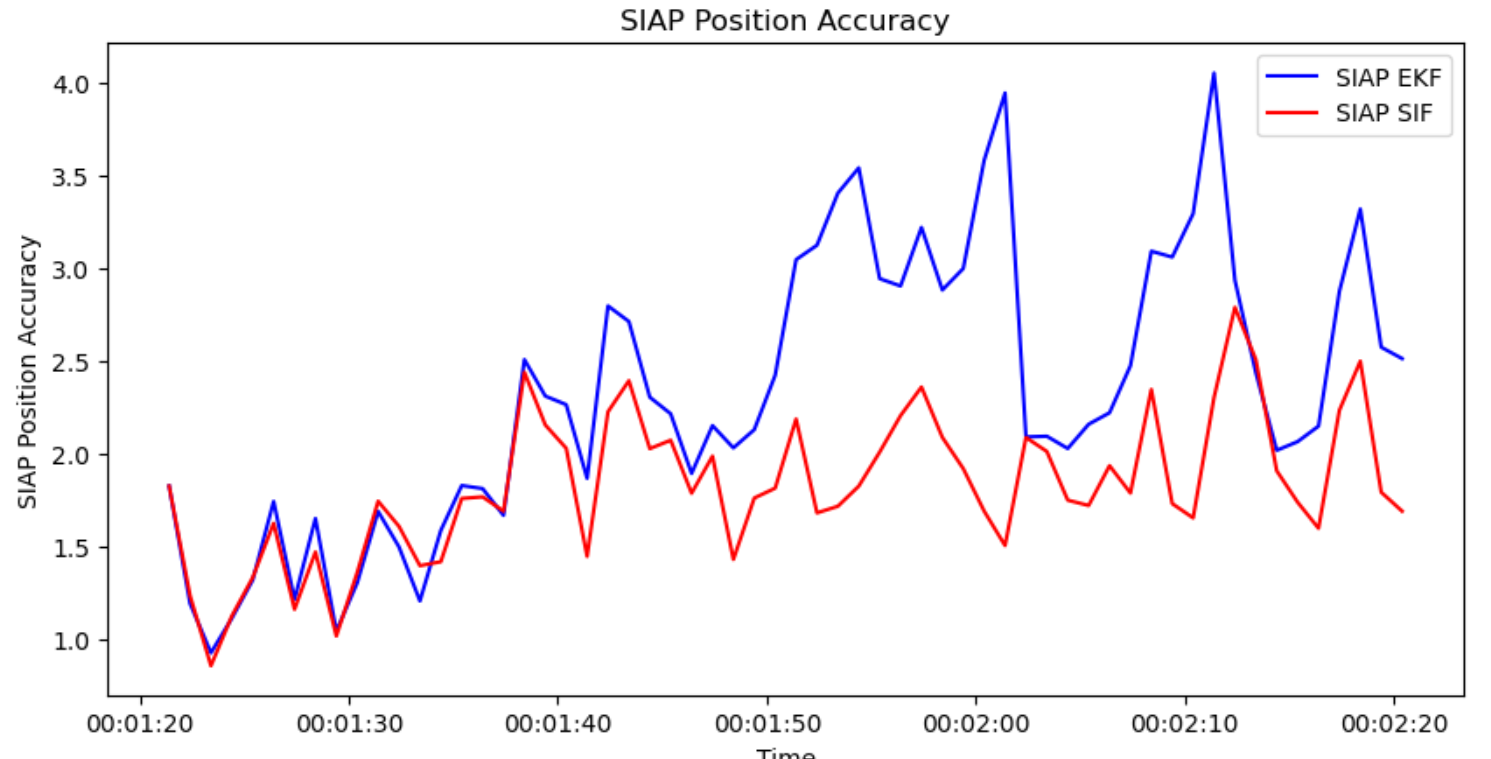}\vspace*{-2mm}
	\caption{SIAP position accuracy results for Class-B example.}
	\label{fig:SIAP1}\vspace*{-4mm}
\end{figure}

\begin{figure}[]
	\centering
	\includegraphics[width=1\linewidth]{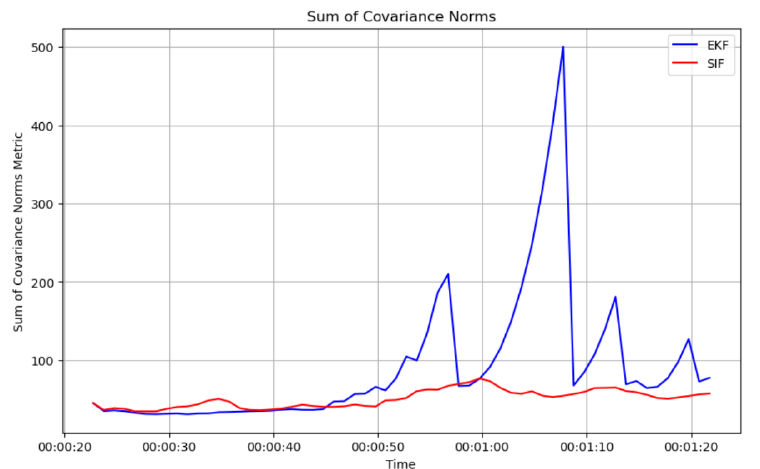}\vspace*{-2mm}
	\caption{Covariance norms metrics results for Class-B example.}
	\label{fig:covNorms}\vspace*{-4mm}
\end{figure}

The SIF and EKF perform extremely similarly initially, before both filters spike in OSPA. The difference in overall shape of Fig. \ref{fig:OSPA1} and Fig. \ref{fig:SIAP1} highlight a key difference in OSPA and positional error as metrics. From time $k=45$ onward, both the positional error and OSPA graphs show the EKF and SIF diverge from one another, but the OSPA graph shows a smaller relative difference between the two trackers. This is due to OSPA emphasizing cardinality errors more than locality errors, thus making the metric less sensitive to being skewed by lone outlier tracks which diverge.

\begin{figure}[]
	\centering
	\includegraphics[width=1\linewidth]{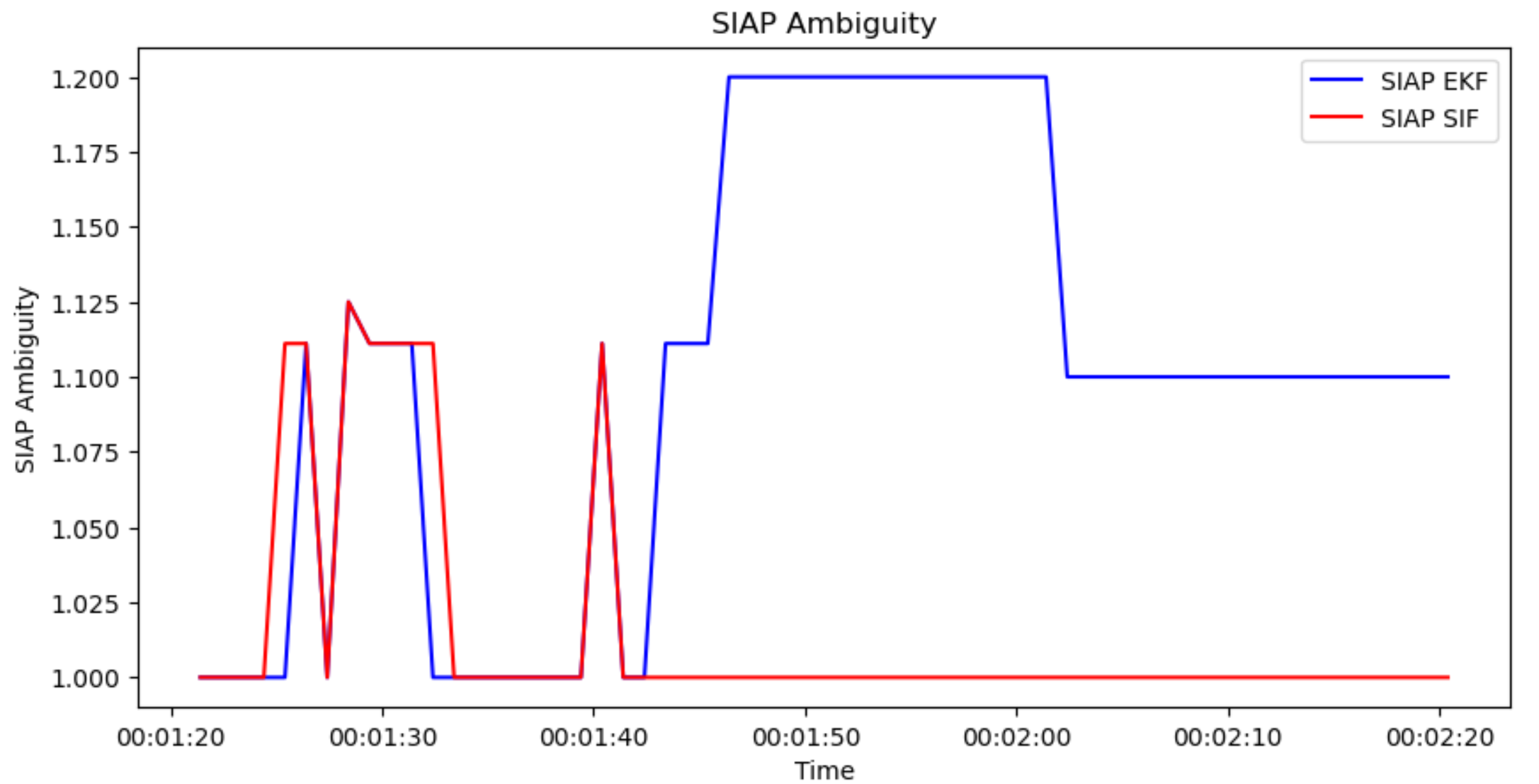}\vspace*{-2mm}
	\caption{SIAP Ambiguity Metric for Class-B example}
	\label{fig:siapambiguity1}\vspace*{-4mm}
\end{figure}

The positional error graph in Fig. \ref{fig:SIAP1} shows that neither filter ever fully diverged or lost any of the targets. But the sum of covariance norms in Fig. \ref{fig:covNorms} indicates that on several occasions the EKF saw massive spikes in uncertainty which are not present for the SIF. These spikes are due to association errors in which the EKF incorrectly distinguishes between two or more nearby targets. This can be seen in Fig. \ref{fig:siapambiguity1} where the EKF fails to distinguish between nearby associations, leading to a spike in SIAP Ambiguity. Interestingly, the ambiguity for the EKF is never fully resolved with the EKF, and this illustrates the interdependent nature of spiking covariance in estimation with errors in association. In this example, the SIF's advantages in covariance propagation enable it to distinguish between targets leading to a much more accurate set of tracks.

\section{OpenSky ADS-B Class-A Airspace Dataset}

ADS-B is a standard aircraft surveillance technology in which aircraft determine their own position and heading via satellite navigation and broadcast it out at set intervals to air traffic control towers and other aircraft in the vicinity. The datasets gathered from aircraft ADS-B systems are freely available to the public and offer an easily accessible and realistic environment to test multi target tracking methods, e.g., via OpenSky Network \cite{SchStLe:14}. However, the ADS-B data has some limitations for the purpose of Air Traffic Control such as (1) antenna positions, (2) bandwidth variations, (3) local policies, (4) data repository, and (5) interval rates. Hence, research challenges include coverage, GNSS validation, and decoding. These limitations along with the vulnerability of ADS-B to cyberattacks via spoofing false aircraft or jamming the broadcasts of true ones, necessitate MTT methods to ensure the integrity of the surveillance system.

% ----
% In these two scenarios, trackers are constructed from two identically configured systems of a data associator, track initiator, an estimator, and a covariance based track deleter to terminate tracks.
% -----

\subsection{Simulating Detections}
In this example, the position and velocity data from aircraft in Great Britain is treated as the ground truth. Detections are simulated from two stationary points which broadly correspond to ground based radars located in Manchester airport and Heathrow airport in London. A third sensor is located on an airborne moving platform located with initial state at longitude and latitude $(52.25\si{\degree},-0.09\si{\degree})$, at an elevation $5\ [km]$ above sea level, which is between the cities of Manchester and London. These coordinates were chosen specifically to deal with the finite range of the sensors \cite{We:00}. %, which have a maximum range of $111\ [km]$ 
The detections are converted from longitude and latitude, into bearing range, and angle of elevation format to simulate radar detections in three dimensions. A total of 84 aircraft are detected by the simulated sensors and are tracked for as long as they remain in range of the sensors.

% The aircraft are tracked by three radars providing measurement of the bearing, azimuth, and range with the zero-mean additive noise with the covariance matrix.
% \begin{align*}
%     R = \begin{bmatrix}
%     \sigma_\theta^2 & 0 & 0 \\
%     0 & \sigma_\phi^2 & 0 \\
%     0 & 0 & \sigma_r^2
%     \end{bmatrix} = \begin{bmatrix}
%     (0.75\pi/180)^2 & 0 & 0 \\
%     0 & (2\pi/180)^2 & 0 \\
%     0 & 0 & 100^2
%     \end{bmatrix}.
% \end{align*}

% The maximum range of $111 [km]$ and the covariance parameters chosen here were picked to resemble the Airport Surveillance Radar 11 (ASR-11) \cite{We:00}. %\textcolor{red}{Could someone find a reference for me? I see a cited reference "FAA SURVEILLANCE RADAR DATA" which appears to outline more exact specifications for the ASR-11, author Mark. E. Weber- I am not sure I understand, what do you mean?}.

% Sensor Parameters

% \begin{align*}
%     R = \begin{bmatrix}
%     \sigma_\theta^2 & 0 & 0 \\
%     0 & \sigma_\phi^2 & 0 \\
%     0 & 0 & \sigma_r^2
%     \end{bmatrix} = \begin{bmatrix}
%     0.75^2 & 0 & 0 \\
%     0 & 2^2 & 0 \\
%     0 & 0 & 100^2
%     \end{bmatrix}
% \end{align*}

\subsection{Tracker Parameters}
In this scenario we expand the simulation from two dimensions via bearing and range, to three by considering the azimuth, elevation, and range model to expand the scenario to three dimensions. This scenario also involves a significantly higher number of aircraft. Due to the dataset considering motion in three dimensions, the transition model for estimating the velocity of aircraft changes and we consider the constant velocity model in three dimensions, with corresponding covariance matrix:
\begin{align*}
    \bfF_k &= \begin{bmatrix}
        1 & d t & 0 & 0& 0 & 0\\
        0 & 1 & 0 & 0 & 0 & 0\\
        0 & 0 & 1 & dt & 0 & 0 \\
        0 & 0 & 0 & 1& 0 & 0 \\
        0 & 0 & 0 & 0 & 1 & dt \\
        0 & 0 & 0 & 0 & 0 & 1
    \end{bmatrix},\\
     \bfQ_k &= \begin{bmatrix}
        q_x\Sigma & \bfnul_2 & \bfnul_2\\
        \bfnul_2 & q_y\Sigma &  \bfnul_2\\
        \bfnul_2 & \bfnul_2& q_z\Sigma & \\
        \end{bmatrix},
\end{align*}
with $q_x = 10$, $q_y = 10$, and $q_z = 5$.  The aircraft are tracked by three radars providing measurement of the bearing, azimuth, and range.
\begin{align}
    \bfh_k(\bfx_k)=\left[\begin{matrix}
        \arcsin(\frac{\bfx_3}{\sqrt{(\bfx_{1,k}-r_\mathrm{N})^2+(\bfx_{3,k}-r_\mathrm{E})^2+(\bfx_{5,k}-r_\mathrm{D})^2}}) \\
        \arctan\tfrac{\bfx_{3,k}^2-r_\mathrm{E}}{\bfx_{1,k}^2-r_\mathrm{N}} \\
        \sqrt{(\bfx_{1,k}-r_\mathrm{N})^2+(\bfx_{3,k}-r_\mathrm{E})^2+(\bfx_{5,k}-r_\mathrm{D})^2} \\
    \end{matrix}\right],
\end{align}
with the zero-mean additive noise with the covariance matrix
\begin{align*}
    \bfR_k = \begin{bmatrix}
    \sigma_\theta^2 & 0 & 0 \\
    0 & \sigma_\phi^2 & 0 \\
    0 & 0 & \sigma_r^2
    \end{bmatrix} = \begin{bmatrix}
    (0.75\pi/180)^2 & 0 & 0 \\
    0 & (2\pi/180)^2 & 0 \\
    0 & 0 & 100^2
    \end{bmatrix}.
\end{align*}
The radar position $[r_\mathrm{N},r_\mathrm{E}, r_\mathrm{D}]^T$ in three dimensional space which is assumed known for all three radars. This includes the radar on the moving platform. 
%The measurement is corrupted by the measurement noise $\bfv_k$ with the covariance matrix.
The maximum range of $111\ [km]$ and the covariance parameters chosen here were picked to resemble the Airport Surveillance Radar 11 (ASR-11) \cite{We:00}.

% Likewise the aircraft are tracked by three radars providing measurement of the bearing, azimuth, and range with the zero-mean additive noise with the covariance matrix.

% \begin{align}
%     \bfh_k(\bfx_k)=\left[\begin{matrix}
%         \arcsin(\frac{\bfx_3}{\sqrt{(\bfx_{1,k}-r_\mathrm{N})^2+(\bfx_{3,k}-r_\mathrm{E})^2+(\bfx_{5,k}-r_\mathrm{D})^2}}) \\
%         \arctan\tfrac{\bfx_{3,k}^2-r_\mathrm{E}}{\bfx_{1,k}^2-r_\mathrm{N}} \\
%         \sqrt{(\bfx_{1,k}-r_\mathrm{N})^2+(\bfx_{3,k}-r_\mathrm{E})^2+(\bfx_{5,k}-r_\mathrm{D})^2} \\
%     \end{matrix}\right],
% \end{align}
%\textcolor{red}{Could someone fix this equation being a tiny bit too wide? - It seems allright it does not go over the text.}
%,r_\mathrm{E}, r_\mathrm{D}]^T$ denotes the position of the sensor in 3 dimensional space which is assumed known for all three radars. This includes the radar on the moving platform. The measurement is corrupted by the measurement noise $\bfv_k$ with the covariance matrix.

\subsection{Results}

\begin{figure}[]
	\centering
	\includegraphics[width=1\linewidth]{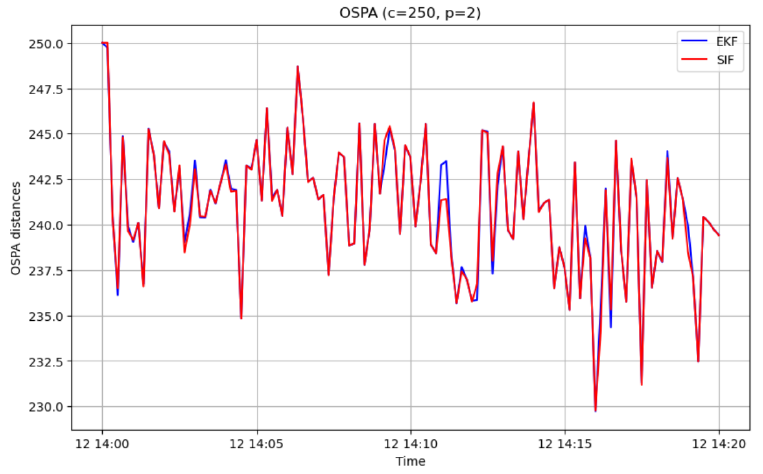}\vspace*{-2mm}
	\caption{OSPA metrics results for ADS-B Dataset}
	\label{fig:OSPA2}\vspace*{-4mm}
\end{figure}

\begin{figure}[]
	\centering
	\includegraphics[width=1\linewidth]{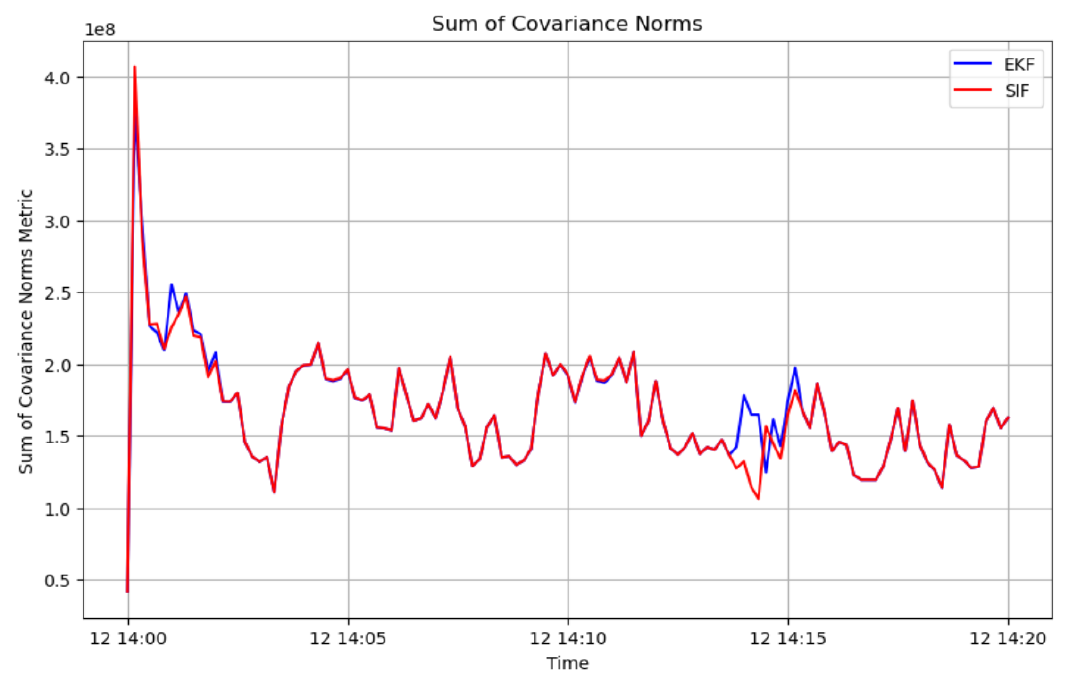}\vspace*{-2mm}
	\caption{Covariance norms metrics result for the ADS-B Dataset}
	\label{fig:covNorms2}\vspace*{-4mm}
\end{figure}

\begin{figure}[]
	\centering
	\includegraphics[width=1\linewidth]{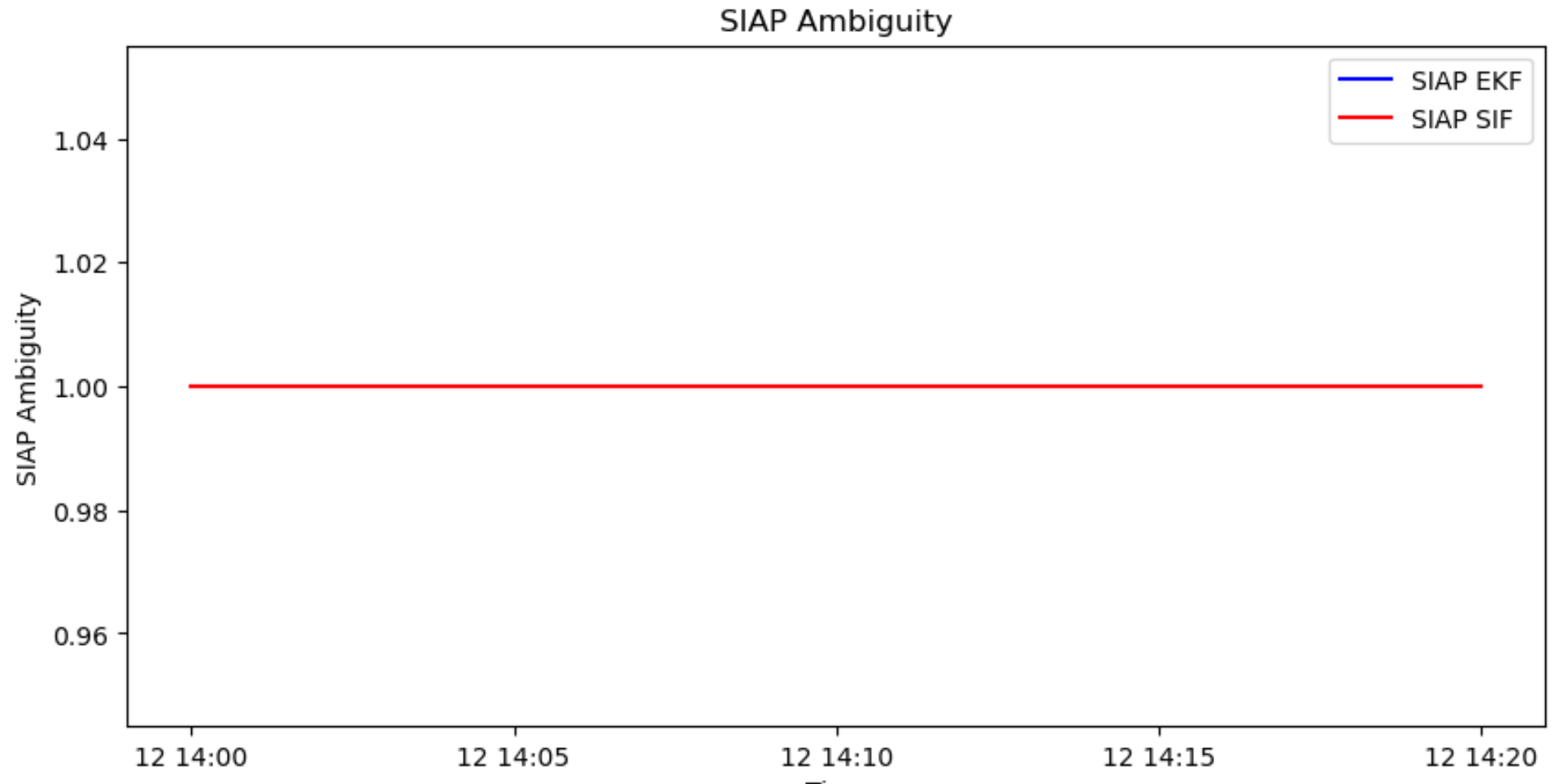}\vspace*{-2mm}
	\caption{SIAP Ambiguity Metric for Class-A example}
	\label{fig:SIAP Ambiguity Ex2}\vspace*{-4mm}
\end{figure}

For the Class-A airspace example, the SIF demonstrated a marginal improvement over the EKF. At first glance a reasonable sounding explanation is that targets in this scenario do not perform maneuvers very often, as aircraft tend to ascend to cruising altitude where they travel at constant speed and heading. However, if model mismatch due to maneuvering targets were the dominant factor in the spiking error, then Fig. \ref{fig:covNorms} would show corresponding spikes in covariance in the SIF curve which are not present in the Class-B Airspace example. This indicates that the similar performance of the two algorithms is instead explained by the fact that targets in Class-A airspace are so much farther apart, meaning that target ambiguity is less of a concern. This can most evident in Fig. \ref{fig:SIAP Ambiguity Ex2} which shows that none of the filters falsely associate detections. In this scenario the linearization of the EKF offers a sufficient approximation of the state uncertainty such that the benefits of the SIF do not prove decisive.

\section{Conclusions}

In both examples the third order SIF demonstrated a notable albeit subtle performance uplift over the Extended Kalman Filter. While the differences were much smaller in the Class-A scenario, the SIF demonstrated a marginal improvement in covariance propagation which in turn led to a more consistent tracker. Where maximum performance is desired, the SIF offers a significant advantage in scenarios where the increased computational complexity of the SIF compared to the EKF is not a constraint.
The SIF demonstrates particular promise in high density spaces with a large amounts of targets per zone where sensors may have trouble associating tracks to targets. This implies that the SIF may have a significant advantage in tracking targets in the presence of clutter, though this scenario was not tested in this paper and will be saved for follow up studies.

% \section*{Acknowledgment}

\bibliographystyle{splncs_srt}

%\vspace{12pt}
%\color{red}
%IEEE conference templates contain guidance text for composing and formatting conference papers. Please ensure that all template text is removed from your conference paper prior to submission to the conference. Failure to remove the template text from your paper may result in your paper not being published.

\end{document}